\documentclass[aps,prb,preprint]{revtex4}
\usepackage{graphicx,psfrag}
\usepackage{graphicx} 
\usepackage{bm}
\usepackage{booktabs}
\usepackage{amsmath,amsthm,amssymb}
\newlength{\minitwocolumn}
\begin{document}

\title{Theory of inverse Faraday effect in disordered metal  in THz regime}

\author{Katsuhisa Taguchi and  Gen Tatara }  

\affiliation{%
Department of Physics, Tokyo Metropolitan University, 
 Hachioji, Tokyo 192-0397, Japan
}

\date{\today}
%
\begin {abstract} 

We calculate the magnetization dynamics induced by the inverse Faraday effect in disordered metals in THz regime by using the diagrammatic method. 
We find that the induced magnetization is proportional to the frequency of circularly polarized light.
\end{abstract}

\maketitle

\section{Introduction} 
Photoinduced magnetization by circularly polarized light (the inverse Faraday effect) has been studied since its theoretical prediction in the 1960s \cite{rf:Pitaevskii61,rf:Pershan63,rf:Ziel65, rf:Pershan66, rf:Hertel06}. 
Recently, experiments using intense laser have been conducted, and the induced effective magnetic field was shown to be as strong as a few Tesla \cite{rf:Kimel05,rf:Stanciu07,rf:Kirilyuk,rf:Hansteen05}.
The fast magnetization reversal realized is expected to be useful for application to magnetic devices \cite{rf:Hansteen05}.

The mechanism of the inverse Faraday effect was discussed by Pitaevskii based on a symmetry argument \cite{rf:Pitaevskii61}.
He noted that the free energy of the electromagnetic field in solids is generally given by 
\begin{align}
F=\sum_{ij}{\rm Re} (\varepsilon_{ij}E_i E_j^*),
\end{align}
where $\varepsilon_{ij}$ is a $3\times3$ tensor of the dielectric functions dependent on mediums, and $E_i$ is the complex representation of the electric field.
The magnetization induced by light is given by the functional derivative of $F$ with respect to the external magnetic field, $H^{\rm{ex}}$, as
\begin{align}
M_k=\sum_{ij}{\rm Re} \left(\frac{\partial \varepsilon_{ij}}{\partial H^{\rm{ex}}_k} E_i E_j^*\right)|_{H^{\rm{ex}}=0}.
\end{align}
Because of the Onsager relation, the asymmetric components of $\varepsilon_{ij}$ are linear or higher odd power in the external magnetic field, while the symmetric components are an even power of the magnetic field  \cite{rf:Pitaevskii61, rf:Landau}.
Therefore, expanding up to the linear order in the magnetic field, the dielectric tensor is 
$\varepsilon_{ij}=\varepsilon_0\delta_{ij}+i\chi_1\epsilon_{ijk}H^{\rm{ex}}_k$, where  
$\varepsilon_0$ and $\chi_1$ are constants independent of $H^{\rm{ex}}$.
The induced magnetization is therefore written generally as
\begin{align}
\bm{M} = i \chi_1 \bm{E}\times \bm{E}^*. 
\end{align}
It is thus proportional to the helicity vector of the circularly polarized light, $i(\bm{E}\times \bm{E}^*)$.
($ \bm{E} \times \bm{E}^*$ is parallel to light propagation and is pure imaginary)
\cite{rf:Pitaevskii61,rf:Pershan63,rf:Ziel65, rf:Pershan66, rf:Hertel06}.
Because the helicity vector, $i(\bm{E}\times \bm{E}^*)$, of the oscillate phase factor is canceled out for $Ee^{i\Omega t}$ and $ E^* e^{-i\Omega t}$, $M$ is a static and rectified magnetization.
This magnetization is converted from the optical angular momentum of circularly polarized light, $\bm{E}\times \bm{E}^*$, and this conversion efficiency is determined by $\chi_1$. 
The coefficient $\chi_1 $ was explicitly calculated in the visible light case considering plasma oscillation \cite{rf:Hertel06}.  It was found to strongly depend on the angular frequency $\Omega$ as 
$\chi_1 \propto \Omega^{-3}$. 
Another mechanism of the inverse Faraday effect is spin-orbit interaction,  as discussed by Pershan and coworkers \cite{rf:Pershan63, rf:Ziel65, rf:Pershan66}. They also showed that the spin-orbit contribution also results in $\chi_1 \propto \Omega^{-3}$. 

Besides the effect of the visible light lasers, the effect of terahertz lasers on magnetic systems has recently been attracting interest.
For example, the spin structure and excitation in ferroelectric magnets were explored by utilizing use of the magnetoelectric effect due to a terahertz laser \cite{rf:kida08}. 

In this paper, we will theoretically investigate the magnetization induced by the inverse Faraday effect in a THz regime.
The magnetization is calculated using the diagrammatic method in disordered metals considering the spin-orbit interaction. The induced magnetization is found to linearly depend on the frequency in the THz regime. 

\section{Hamiltonian}
We consider the conduction electrons in the presence of spin-orbit interaction and impurity scattering.
The applied THz light is represented by 
the vector potential $\bm{A}_{\rm{em}}$. 
The Hamiltonian of the system is  
$\mathcal{H}\equiv \mathcal{H}_0 + \mathcal{H}_{\rm{em}} + \mathcal{H}_{\rm{so}} + \mathcal{H}_{\rm{i}}
$, where 
\begin{align}
\mathcal{H}_0
& = \int d\bm{x} \frac{\hbar^2}{2m} \bm{\nabla}  c^\dagger \cdot \bm{\nabla} c,
\\
\mathcal{H}_{\rm{em}}
	& = -\int d\bm{x}   c^\dagger \frac{e }{2m}  (\bm{p}\cdot \bm{A}_{\rm{em}}+ \bm{A}_{\rm{em}} \cdot\bm{p} -eA^2_{\rm{em}} )c , 
\\
\mathcal{H}_{\rm{so}} & = - \frac{\lambda}{\hbar} \int d\bm{x}^3 c^\dagger \bm{\sigma}\cdot 
[	( e\bm{\dot{A}}_{\rm{em}}	-  \bm{\nabla}u_{\rm{i}}) \times (\bm{p}-e\bm{A}_{\rm{em}} )] c, 
\\
\mathcal{H}_{\rm{i}}
& = \int d\bm{x} u_{\rm{i}}  c^\dagger c,
\end{align}
and
$c$,  $c^\dagger$ are the conduction electron's annihilation and creation operators, respectively.  
We consider two origins for spin-orbit interaction: one from the applied electromagnetic field and the other induced by  random impurities.
The coefficient $\lambda  \equiv \frac{ \hbar^2 } {4m^2 c^2 } $ is the strength of spin-orbit coupling,  and $u_{\rm{i}} = u \sum_{\ell} \delta(\bm{x} - \bm{R}_\ell)$ is the impurity potential, where $\bm{R}_\ell$ represents the randomly distribution impurity positions.
(The average of $\lambda u$ represents the strength of the spin-orbit interaction in the material.)
The scattering of the electron by the impurities  is  represented by $\mathcal{H}_{\rm{i}}$.  We consider the elastic lifetime  of the electron, $\tau$, arising from this scattering.
 
We consider a monochromatic light with a frequency $\Omega$ and wave vector $\bm{q}$(The unit vector of electromagnetic field, $\hat{\bm{q}}=\bm{q}/|q|$, is parallel to $\bm{E}\times \bm{E}^*$).
The vector potential is related to the electric field by  $\bm{\mathcal{E}}_{\rm{em}}= - \dot{\bm{A}}_{\rm{em}}$.
We express the vector potential and electric field by utilizing the complex amplitudes $\bm{E}$ and $\bm{A}$ as 
\begin{align}\label{eq:e1}
\bm{\mathcal{E}}_{\rm{em}} & = 
	\frac{1}{2} (\bm{E} e^{i(\bm{q}\cdot\bm{x}-\Omega t)} +\bm{E}^* e^{-i(\bm{q}\cdot\bm{x}-\Omega t)}), 
\\ \label{eq:a1}
\bm{A}_{\rm{em}}&= \frac{1}{2} ( \bm{A} e^{i(\bm{q}\cdot\bm{x}-\Omega t)} 
	+\bm{A}^* e^{-i(\bm{q}\cdot\bm{x}-\Omega t)}).
\end{align}
The amplitude thus satisfies 
$\bm{A}=-i\frac{\bm{E}}{\Omega}$ and $\bm{A}^*=i\frac{\bm{E}^*}{\Omega}$.

\section{Calculation of the magnetization due to the inverse Faraday effect}

The spin density induced by the applied circularly polarized light (inverse Faraday effect) is calculated by estimating the expectation value  $s^\alpha\equiv \langle c^\dagger \sigma^\alpha c  \rangle$ ($\alpha = x, y, z$ represents the spin direction, and $\langle \ \rangle$ denotes expectation values).
Using the lesser Green's function, $G^<(x, x; t, t) $, the spin density is expressed  
\begin{align}
s^\alpha (\bm{x}, t) &=- i\hbar {\rm{Tr}} \left[ \sigma^\alpha G^<(\bm{x}, \bm{x}; t, t) \right].
\end{align}
The calculation of the spin density is performed by treating $\mathcal{H}_{\rm{em}}$ and $\mathcal{H}_{\rm{so}}$ perturbatively.
From the spin density, the magnetization of inverse Faraday effect is given by  
$ M^\alpha =- \frac{g\mu_{\rm{B}}}{2} s^\alpha$, 
where 
$\mu_{\rm{B}}$ is the Bohr magneton, and 
${\it{g}}$ is the  ${\it{g}}$-factor. 
In the following section, we only consider the contribution proportional to ${\bm E}\times {\bm E}^*$, because we are interested in the magnetization induced by the inverse Faraday effect.
\subsection{Spin-orbit interaction from the electric field}
We first estimate the spin-orbit interaction arising from the applied electric field, whose Hamiltonian  we represent as 
$\mathcal{H}_{\rm{so}}^{(1)}$, i.e.,
\begin{align}
\mathcal{H}_{\rm{so}}^{(1)} & \equiv  - \frac{\lambda}{\hbar} \int d\bm{x}^3 c^\dagger \bm{\sigma}\cdot 
[	 e\bm{\dot{A}}_{\rm{em}} \times (\bm{p}-e\bm{A}_{\rm{em}} )] c. 
\end{align}
The contribution to the spin density, $s^{(1)\alpha}$, due to $\mathcal{H}_{\rm{so}}^{(1)}$ are shown in Fig. 1, and are calculated as 
\begin{align} \label{eq:s-11}
s^{(1) \alpha} &= - i2e^2\lambda \epsilon_{\alpha \gamma \ell }
	\left(E_\gamma A^*_\ell  + E_\gamma^* A_\ell  \right)
K(\Omega),
\end{align}
where 
\begin{align}
 K(\Omega) & \equiv \sum_{k,\omega}\left[  \left[ g_{k, \omega}\ g_{k, \omega}  \right]^<
+	\frac{\hbar^2 }{m}  \left[ I_{\ell \zeta} (\Omega)+I_{\ell \zeta}(-\Omega)  \right] \right], 
\\ \notag
I_{\ell \zeta} (\Omega) & \equiv \frac{\hbar^2}{m} \sum_{k,\omega} 
		k_{\ell} k_\zeta 	\left[ g_{k, \omega}\ g_{k, \omega+\Omega} \ g_{k, \omega}  \right]^<, 
\end{align}
and $g_{k,\omega}$ is the free Green's function on a Keldysh contour.
The lesser component of the free Green's function  is 
$g_{k,\omega}^< = f(\omega)(g_{k,\omega}^a -g_{k,\omega}^r)$, 
where $f(\omega)$ is the Fermi distribution function, $g^a = (\hbar\omega -\epsilon_k -i\frac{\hbar}{2\tau})^{-1}$ and $g^r= (g^a)^*$ are the advanced and retarded Green's functions, respectively\cite{rf:tatara08}. 
($\epsilon_k \equiv \frac{\hbar^2 k^2}{2m} - \epsilon_{\rm{F}}$ and $\epsilon_{\rm{F}}$ is the Fermi energy).
Here $K(\Omega)$ is the even function with respect to $\Omega$. 
\begin{figure}\centering
\includegraphics[scale=0.35]{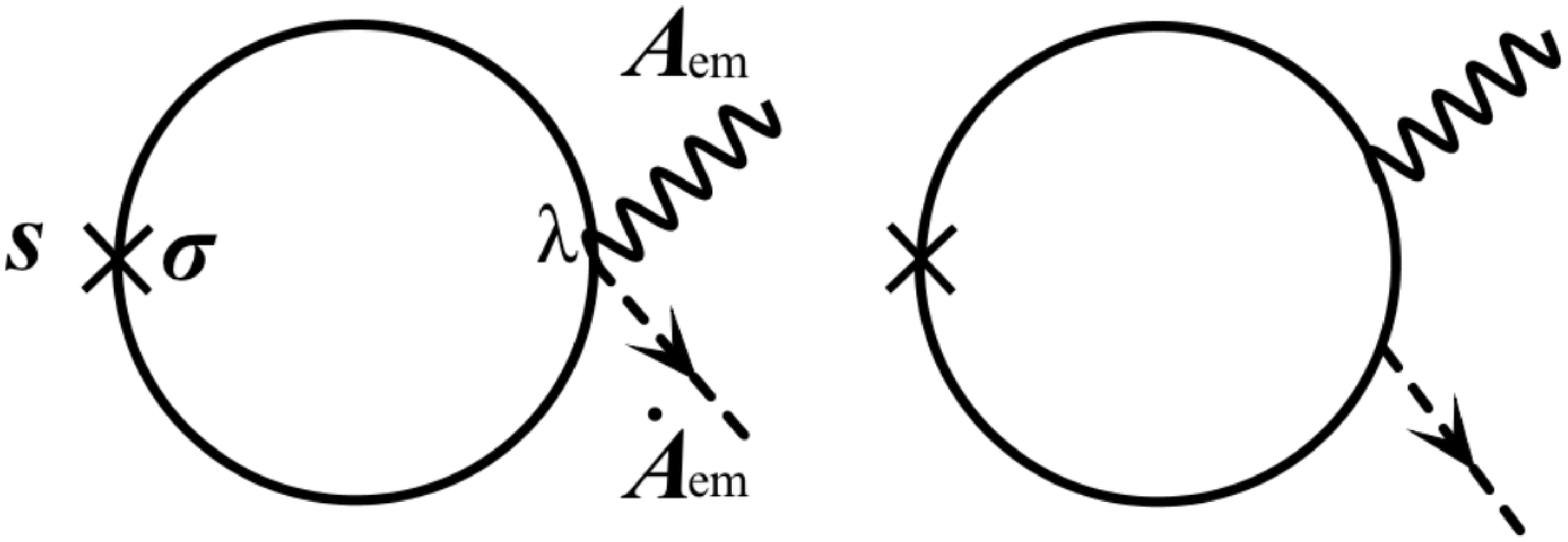}
\caption{
Diagrams contributing to the spin density $\bm{s}^{(1)}$ induced by circularly polarized light.  The electron's Green's functions including the impurity average here are denoted by thick lines. 
The wavy line and the dashed arrow represent the gauge field of light ($\bm{A_{\rm{em}}}$) and  electric field ($ \dot{\bm{A}}_{\rm{em}}$), respectively. 
$\lambda$ is spin-orbit coupling constant accompanied with $ \dot{\bm{A}}_{\rm{em}}$.
The vertex shows the Pauli matrix ($\sigma$).   	\label{Fig:1}}
\end{figure}

In the terahertz regime, 
the frequency of light, $ \Omega$, is much smaller than that of the electron, $\omega$.
We therefore expand the Fermi distribution function as
$f(\omega+ \Omega) = f(\omega) + \Omega f'(\omega) + \frac{\ \Omega^2}{2} f''(\omega) + o(\Omega^3)$.
We thus obtain
\begin{align}\notag
K(\Omega)
=&\sum_{k,\omega} \frac{\hbar^2 k_{\ell} k_\zeta }{2m} f'_\omega \hbar\Omega^2 
	 ((g_{k,\omega}^a )^4 -(g_{k,\omega}^r )^4 )  +o(\Omega^4),
\end{align}
where we have used the partial integration with respect to $k$ 
(Note that $K(0) = 0$).
Considering zero temperature, i.e., $f'(\omega) = - \delta(\omega)$, we obtain $K(\Omega)$ as
\begin{align}\label{eq:pure-3}
K(\Omega) = \frac{i\pi \nu} {24 \epsilon_{\rm{F} }^2 }\hbar\Omega^2 \delta_{\ell \zeta}  +o(\Omega^4).
\end{align}
From  Eqs. (\ref{eq:s-11}) and (\ref{eq:pure-3}), 
the result of the induced spin density is given by  
\begin{align}\label{eq:spin density-so}
s^{(1)\alpha} & = i \frac{\pi \nu e^2 \lambda} {6 \epsilon_{\rm{F}}^2 }\hbar\Omega
\epsilon_{\alpha \beta \gamma } E_\beta E_\gamma^*,
\end{align}
where $\nu$ is the electron density of state at the Fermi energy per volume.

\subsection{Spin-orbit interaction due to impurities}
Next, we consider the inverse Faraday effect from spin-orbit interaction caused by the random impurity potential: 
\begin{align}
\mathcal{H}_{\rm{so}}^{(2)} & \equiv 
 \frac{\lambda}{\hbar} \int d\bm{x}^3 c^\dagger \bm{\sigma}\cdot 
[  \bm{\nabla}u_{\rm{i}} \times (\bm{p}-e\bm{A}_{\rm{em}} )] c.
\end{align}
The random potential is treated by averaging,   
$\langle u_{\rm{i}}(q) u_{\rm{i}}(q') \rangle_{\rm{i}}= n_{\rm{i}} u_{\rm{i}}^2 \delta(\bm{q}+\bm{q}')$ and 
$\langle u_{\rm{i}}(q) u_{\rm{i}}(q') u_{\rm{i}}(q'') \rangle_{\rm{i}}= n_{\rm{i}} u_{\rm{i}}^3 \delta(\bm{q}+\bm{q}'+ \bm{q}'')$, where $\bm{q}$, $\bm{q}'$, $\bm{q}''$ are wave numbers, $n_{\rm{i}}$ is the concentration of the impurity,  and $\langle \ \rangle_{\rm{i}}$ is the random impurity average.

\begin{figure}\centering
\includegraphics[scale=0.4]{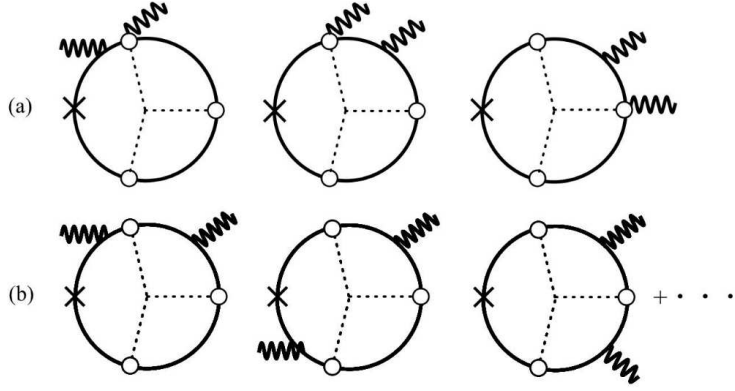}
\caption{ \label{Fig:2}
Diagrams contributing to the spin density in the spin-orbit interaction due to random impurities.
Dashed lines denote the impurity averaging between the spin-orbit interaction¡¡(the combination between the wavy line and the open circle)  and the impurity scattering (open circle).
The contribution of diagrams (a) and (b) indicates $s^{(2)\alpha}_{3(a)}$ and $s^{(2)\alpha}_{3(b)}$, respectively.}
\end{figure}

\begin{figure}\centering
\includegraphics[scale=0.35]{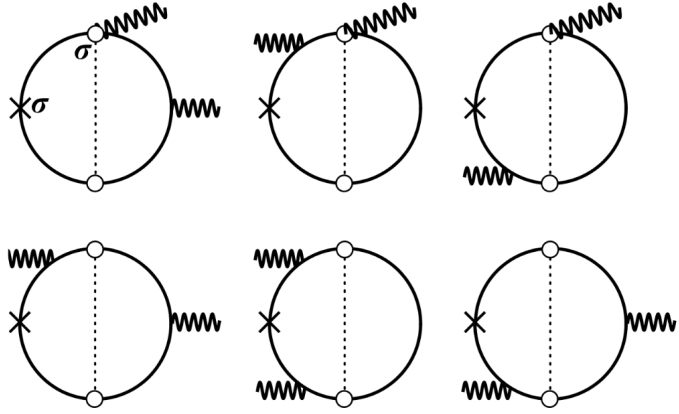}
\caption{ \label{Fig:3}
Diagrammatic representation of  spin density.	}
\end{figure}

Contributions to the spin density proportional to ${\bm E}\times {\bm E}^*$ are diagrammatically shown in Figs. $\ref{Fig:2}$ and $\ref{Fig:3}$.
Processes shown in Fig. $\ref{Fig:2}$ and Fig. $\ref{Fig:3}$ differ in impurity averaging; the averaging in Fig. $\ref{Fig:2}$ is third order in the impurities as in the case of the anomalous Hall effect \cite{rf:Crepieux01}.
We first estimate the contribution (a) of Fig. $\ref{Fig:2}$, defined as $s^{(2)\alpha}_{3(a)}$. It is
\begin{align}
s^{(2)\alpha}_{3(a)} &=  -\frac{2e^2\hbar \lambda nu^3}{3m}  \epsilon_{\alpha \ell \gamma}
 E_\ell E^*_\gamma ( \Gamma _{\Omega} - \Gamma _{-\Omega}).
\end{align}
In the above equation, we have used the relation, $\bm{A}_{\rm{em}}=\frac{1}{2}\left( \frac{\bm{E}}{i\Omega}e^{-i\Omega t} -  \frac{\bm{E}^*}{i\Omega} e^{i\Omega t} \right) $. $\Gamma _{\Omega}$ is written as 
\begin{align}\notag
 \Gamma _{\Omega} &= \frac{2}{\Omega^2}\sum_{k,k',k'',\omega} k^2 (f_{\omega +\Omega} - f_{\omega})
	\begin{pmatrix}
g_{k, \omega}^r g_{k,\omega}^a \ (g_{k, \omega + \Omega}^a  g_{k', \omega +\Omega}^{a}\  g_{k'', \omega +\Omega}^{a} -c.c)\\
-  g_{k, \omega + \Omega}^r\  g_{k, \omega +\Omega}^a  (g_{k,\omega }^a \ g_{k',\omega }^{a}\  g_{k'',\omega }^{a} -c.c)
	\end{pmatrix}
 +\delta \Gamma_\Omega \\ \label{eq:gamma} 
&= \frac{2}{\Omega^2}\sum_{k,k',k'',\omega} k^2 
	\left[\begin{pmatrix}
(f_{\omega +\Omega} - f_{\omega}) 
g_{k, \omega}^r g_{k,\omega}^a \ (g_{k, \omega + \Omega}^a  g_{k', \omega +\Omega}^{a}\  g_{k'', \omega +\Omega}^{a} -c.c) \\
	\end{pmatrix} +\left( \Omega \to -\Omega\right)\right]
 +\delta \Gamma_\Omega,
\end{align}
and $\delta\Gamma$ is given by 
\begin{align} \label{eq:gamma-2}
\delta \Gamma_\Omega &=  \frac{2}{\Omega^2}\sum_{k,k',k'',\omega} k^2 f_\omega \left[
\left(g_{k, \omega}^a g_{k, \omega +\Omega}^a \ ( g_{k', \omega}^{a}\  g_{k'', \omega}^{a} (g_{k, \omega}^a + g_{k', \omega}^{a} +  g_{k'', \omega}^{a}  )
- g_{k, \omega}^a  g_{k', \omega +\Omega}^a \  g_{k'', \omega +\Omega}^a ) \right) -c.c
\right].
\end{align}
The first term of $\Gamma_\Omega$ is even function respect with $\Omega$, and
therefore $ \Gamma _{\Omega} - \Gamma _{-\Omega} = \delta \Gamma_\Omega - \delta \Gamma_{-\Omega}$.
The Green's function can expanded as $g_{k, \omega +\Omega}^a = g_{k, \omega}^a - \hbar\Omega (g_{k, \omega}^a)^2 +  (\hbar\Omega)^2 (g_{k, \omega}^a)^2 - (\hbar\Omega)^3 (g_{k, \omega}^a)^3 +o(\Omega^4)$, and thus
$\delta \Gamma_\Omega - \delta \Gamma_{-\Omega}$ is reduced to
\begin{align}\notag
\delta \Gamma_\Omega - \delta \Gamma_{-\Omega} 
& = 4\hbar^3 \Omega \sum_{k,k',k'',\omega} k^2 f_\omega  
 	\left[	 
	 \begin{pmatrix} 
  +  (g_{k, \omega}^a)^3   (g_{k', \omega}^a )^4 \  g_{k'',\omega}^a 
  +  (g_{k, \omega}^a)^3   g_{k', \omega}^a \  ( g_{k'',\omega}^a )^4 \\
  +  (g_{k, \omega}^a)^4   ( g_{k', \omega}^a )^3 \  g_{k'',\omega}^a 
  +  (g_{k, \omega}^a)^4   g_{k', \omega}^a \  ( g_{k'',\omega}^a )^3 \\
  +  (g_{k, \omega}^a)^3   ( g_{k', \omega}^a)^2 \  (g_{k'',\omega}^a)^3
  +  (g_{k, \omega}^a)^3   (g_{k', \omega}^a)^3 \  (g_{k'',\omega}^a)^2 \\
  +  (g_{k, \omega}^a)^4   (g_{k', \omega}^a)^2 \  (g_{k'',\omega}^a )^2
	\end{pmatrix} -c.c \right] 
\\  \label{eq:gamma'} 
& + o(\Omega^3).
\end{align}
This contribution at low frequency contains only the terms of considering of only the retarded or advanced Green's functions, and it turns out to be negligibly  small compared with $J_\Omega$ of Eq. (\ref{eq:3-2}).

The contribution of Fig.  \ref{Fig:2}(b), $s^{(2)\alpha}_{3(b)}$, is 
\begin{align}
s^{(2)\alpha}_{3(b)} &= \frac{2e^2\hbar\lambda nu^3}{3m }  \epsilon_{\alpha \ell m}
E^*_\ell E_m  		(J_{\Omega} - J_{-\Omega}), 
\end{align}
where  
\begin{align}\notag
J_{\Omega} &\equiv \frac{\hbar^2}{3m\Omega^2} \sum_{k,k',k'' \omega} 
	\begin{pmatrix}
		(k')^2 (k'')^2 \left[ g_{k, \omega}\  g_{k'', \omega} \ g_{k'', \omega+\Omega } 
		\ g_{k', \omega+\Omega} \ g_{k', \omega}\  g_{k, \omega}  \right]^< \\
 	+k^2 (k'')^2 \left[ g_{k, \omega}\ g_{k, \omega+\Omega}  \ g_{k'', \omega+\Omega } 
		\ g_{k'', \omega} \ g_{k', \omega}\  g_{k, \omega}  \right]^<  \\ 
	+k^2 (k'')^2 \left[ g_{k, \omega}\ g_{k'', \omega}  \ g_{k'', \omega+\Omega } 
		\ g_{k', \omega+\Omega} \ g_{k, \omega+\Omega} \  g_{k, \omega}  \right]^<  \\
	+k^2 (k')^2 \left[ g_{k, \omega}\ g_{k'', \omega} \ g_{k', \omega} 
		g_{k', \omega+\Omega } 	\ g_{k, \omega+\Omega} \  g_{k, \omega}  \right]^< \\
	+ k^2 (k')^2 \left[ g_{k, \omega}\ g_{k, \omega+\Omega} \ g_{k'', \omega+\Omega } 
		\ g_{k', \omega+\Omega}\ g_{k', \omega}   \  g_{k, \omega}  \right]^<
		\end{pmatrix}.
\end{align}
The dominant contribution of $J_\Omega$ is estimated as
\begin{align} \label{eq:3-2}
J_\Omega  =& \frac{\hbar^2}{3m\Omega^2}\sum_{k,k',k'',\omega}k^2 (k'')^2	(f_{\omega+\Omega}-f_\omega)
\begin{pmatrix}
	|g^r_{k',\omega}|^2 g^a_{k,\omega}	g^r_{k'',\omega}	(g^a_{k,\omega+\Omega} g^a_{k'',\omega+\Omega}- c.c )	
\\ 
	+|g^r_{k,\omega}|^2 (g^a_{k',\omega} g^a_{k'',\omega} + g^r_{k',\omega} g^r_{k'',\omega}	)(g^a_{k,\omega+\Omega} g^a_{k'',\omega+\Omega} - c.c	)
\\ 
	+ |g^r_{k,\omega}|^2 ( g^a_{k'',\omega} +  g^r_{k'', \omega})(g^a_{k,\omega+\Omega} g^a_{k',\omega+\Omega} g^a_{k'',\omega+\Omega} - c.c	) 
\end{pmatrix}		+ \delta J_\Omega,  
\end{align}
%
where
\begin{align}\label{eq:3-2delta}
\delta J_\Omega & \equiv 		\frac{\hbar^2}{3m\Omega^2}\sum_{k,k',k'',\omega}k^2 (k'')^2  f_\omega
	\begin{pmatrix}
			(g^a_{k',\omega})^2 g^a_{k,\omega} g^a_{k'',\omega} g^a_{k,\omega+\Omega} g^a_{k'',\omega+\Omega}	\\ 
	+ 2 (g^a_{k,\omega})^2 g^a_{k',\omega} g^a_{k'',\omega} g^a_{k,\omega+\Omega} g^a_{k'',\omega+\Omega} \\ 
	+ 2 (g^a_{k,\omega})^2  g^a_{k'',\omega} g^a_{k,\omega+\Omega} g^a_{k',\omega+\Omega} g^a_{k'',\omega+\Omega} 
	\end{pmatrix} -c.c .
\end{align}
Because $\delta\Gamma_\Omega$ and $\delta J_\Omega$ have only the contribution only retarded or advanced Green's functions, Eqs. (\ref{eq:gamma-2}) and (\ref{eq:3-2delta}) are both negligibly small compared with Eq. (\ref{eq:3-2}) (by the order of  $\frac{\hbar}{\epsilon_{\rm{F}}\tau}\ll1$, where $\tau\equiv \frac{\hbar}{2\pi \nu n u^2 }$ is a elastic electron lifetime).
From these estimations, the spin density contribution shown in Fig. \ref{Fig:2},  $s_{3}^{(2)\alpha}  = s_{3(a)}^{(2)\alpha}  + s_{3(b)}^{(2)\alpha} $,  is finally obtained as
\begin{align}\notag
s_{3}^{(2)\alpha} 
&\simeq - i3e^2\hbar\lambda n_{\rm{i}} u^3 \epsilon_{\alpha \ell \gamma}\Omega
E^*_\ell E_\gamma 
\  \left[ \sum_{k} |g_{k}^r|^2 \right]^3 
\\ \label{eq: so-i}
& = i 12\pi^2 e^2 \nu^2 \lambda \left(\frac{u \tau }{\hbar}\right) \Omega \tau \epsilon_{\alpha \ell \gamma} E_\ell E^*_\gamma
\end{align}

The contribution of Fig. $\ref{Fig:3}$ is calculated similarly as
\begin{align} \label{eq:s-2}
s_{2}^{(2) \alpha}  &= \frac{e^2\hbar^2\lambda n_{\rm{i}} u^2}{3m}  \epsilon_{\alpha \ell \gamma}E^*_\ell E_\gamma  
		\left[ I_{\Omega} -  I_{-\Omega} \right], 
\end{align}
where 
\begin{align}\notag
I_{\Omega} \equiv & 
  \sum_{k,k' \omega} 
		\frac{1}{\Omega^2}(k')^2  \left[ g_{k, \omega}\  g_{k', \omega} \ g_{k', \omega+\Omega} \ g_{k, \omega} + g_{k, \omega}\   g_{k', \omega+\Omega}\ g_{k', \omega} \ g_{k, \omega}   \right]^<
\\ \notag
+ & \frac{1}{\Omega^2}
		\sum_{k,k' \omega}  k^2  \left[ g_{k, \omega}\  g_{k', \omega}\ g_{k, \omega+\Omega} \ g_{k, \omega}  + g_{k, \omega}\   g_{k, \omega+\Omega}\ g_{k', \omega} \ g_{k, \omega}  \right]^<
\\ 
+ &	\frac{1}{\Omega^2}\sum_{k,k' \omega}  k^2  \left[ g_{k, \omega}\   g_{k', \omega+	\Omega}\ g_{k, \omega+\Omega} \ g_{k, \omega}  +
		 g_{k, \omega}\  g_{k, \omega+\Omega} \ g_{k', \omega+\Omega} \ g_{k, \omega}  \right]^<.		
\end{align}
The odd function of $\Omega$, $I_{\Omega} -  I_{-\Omega} $, is calculated as
\begin{align} \notag
I_{\Omega} -  I_{-\Omega} & =
\frac{2}{\Omega^2} \sum_{k,k' \omega} k^2   
\begin{pmatrix}
	& f_\omega  
			\left[ (g_{k, \omega}^a (g_{k', \omega}^a)^2 \ g_{k, \omega+\Omega}^a
					+(g_{k, \omega}^a)^2 \ g_{k, \omega+\Omega}^a (g_{k', \omega}^a- g_{k', \omega+\Omega}^a) )- c.c
			\right]
\\ \notag
		& + (f_{\omega+\Omega}- f_\omega) 
		g_{k, \omega+\Omega}^a\ g_{k, \omega+\Omega}^r\ 
		 (g_{k, \omega}^a\ g_{k', \omega}^a- g_{k, \omega}^r\ g_{k', \omega}^r)
\\ \notag
		 & - (f_{\omega+\Omega}- f_\omega) 
		 g_{k, \omega}^a\ g_{k, \omega}^r\ 
		 (g_{k, \omega+\Omega}^a\ g_{k', \omega+\Omega}^a- g_{k, \omega+\Omega}^r\ g_{k', \omega+\Omega}^r) 
\end{pmatrix}
\\ \notag
&=4\hbar^3 \Omega  \sum_{k,k' \omega} k^2   
	  f_\omega  
			\left[ (g_{k, \omega}^a)^3\  (g_{k', \omega}^a)^3 \ 
						(g_{k, \omega}^a+ g_{k', \omega}^a)  - c.c
			\right] +o(\Omega^4)
\\ \label{eq:I-2}
&=m \Omega \frac{i\pi^2 \nu^2 \eta}{2\epsilon_F^4}+o(\Omega^3)
\end{align}
From Eqs. (\ref{eq:s-2}) and (\ref{eq:I-2}), $\bm{s}_2^{(2)}$ reads 
\begin{align}\label{eq:so-imp2}
s_{2}^{(2)\alpha}  
& =   \frac{i \pi^2}{12} \frac{ e^2 \lambda \nu \hbar \Omega}{\epsilon^2_{{\rm{F}}}} \left( \frac{\hbar}{\epsilon_{{\rm{F}}}\tau }\right)^2  E_\ell E^*_\gamma,
 \end{align}
Equation  (\ref{eq:so-imp2}) is small compared with  Eq. (\ref{eq: so-i}) 
by the order of 
$ \frac{\epsilon_{\rm{F}}}{u} \left( \frac{\hbar}{ \epsilon_{\rm{F}}  \tau  } \right)^{2} \ll1$ and thus neglected below.
Therefore, the dominant contribution of $\mathcal{H}_{\rm{so}}^{(2)}$ is given by $s_{3}^{(2)}$. 

We summarize the spin density contributed from the spin-orbit interaction $\mathcal{H}^{(1)}$ and $\mathcal{H}^{(2)}$. 
Since $\bm{s}^{(1)}$ are negligible small compared with $\bm{s}_3^{(2)} $ from Eq. (\ref{eq:spin density-so}) and (\ref{eq:so-imp2}),
the total magnetization of the inverse Faraday effect in spin-orbit interaction, $\bm{M} \simeq -\frac{g\mu_{B}}{2} \bm{s}_3^{(2)} $ is obtained as 
\begin{align}\label{eq:IFM-01}
 \bm{M} & \simeq i \chi  \bm{E} \times  \bm{E}^*, 
\end{align}
where
\begin{align} \label{eq:IFM-01-1}
\chi &=- i 6 g \mu_{\rm{B}} \pi^2 e^2 \nu^2 \lambda \left(\frac{u \tau }{\hbar}\right) \Omega \tau 
\end{align}
The magnetization is proportional to the strength of spin-orbit coupling, intensity of light, frequency, and electron's relaxation time. 

\section{Summary and discussion}
We have shown that 
the spin density induced by the inverse Faraday effect in the THz regime is proportional to $\bm{E}\times \bm{E}^*$. 
From the spin density induced by circularly polarized light, the magnitude of the magnetic field generated in the medium, $\bm{B}^{\rm{eff}}$ can be estimated as $\bm{B}^{\rm{eff}} = \mu \bm{M}$, 
where $\mu$ are magnetic susceptibility in the medium.
Here, we chose the magnetic susceptibility in the vacuum.
The magnitude of spin-orbit coupling is $\lambda k_{\rm{F}}^2 = 0.01 -1$ for various materials\cite{rf:maekawa}, and we choose here $\lambda k_{\rm{F}}^2 \approx 0.5 $ considering the case of platinum\cite{rf:maekawa}. 
We choose the amplitude and frequency of applied electromagnetic field as $|E|=10^8 \rm{V/m}$ and $\Omega = 1 \rm{THz}$, respectively.  
In metals with 
$\epsilon_{\rm{F}} \approx 1$ eV,  
$\frac{u}{\epsilon_{\rm{F}}}=0.1$, and
$\frac{\hbar}{\epsilon_{\rm{F}} \tau} \approx 0.01$,  the magnetic field can be estimated by  $  | B^{\rm{eff}}| \approx 10^{-2}\rm{T}$. 
We have theoretically studied the spin density induced by circularly polarized light in THz regime in metals spin-orbit interaction.
The induced spin is proportional to spin-orbit interaction, frequency of applied THz light, and also depends on the electron¡Çs relaxation time.

\section*{Acknowledgments}
This work was supported by a Grant-in-Aid for Scientific Research (B) (Grant No. 22340104) from Japan Society for the Promotion of Science and UK-Japanese Collaboration on Current-Driven Domain Wall Dynamics from JST.
This work is also financially supported by the Japan Society for the Promotion of Science for Young Scientists.

\end{document}